\documentclass{svproc}
\usepackage{url}
\usepackage{makeidx}
\usepackage{graphicx} 
\usepackage{ragged2e}
\usepackage{amsmath,amssymb,bm}
\usepackage{physics}
\usepackage{ragged2e}
\usepackage{geometry}
\usepackage{newtxmath}
\usepackage{xcolor}
\usepackage[bottom]{footmisc}
\usepackage{newtxtext}

\begin{document}
\mainmatter              
\title{Nonlinear Dynamical Regimes of Cosmological Frequency Combs}
\titlerunning{Nonlinear Dynamical Regimes of Cosmological Frequency Combs}  
\author{
\underline{Madhurendra Mishra}\inst{1}$^\dagger$
\and
Oem Trivedi\inst{2}$^\dagger$
\and
Adarsh Ganesan\inst{3,4}$^*$
}

\authorrunning{Madhurendra Mishra et al.}

\institute{
Department of Physics, Sri Guru Tegh Bahadur Khalsa College, University of Delhi, Delhi, India\\
\and
Department of Physics and Astronomy, Vanderbilt University, Nashville, USA\\
\and
Department of Electrical and Electronics Engineering, Birla Institute of Technology and Science, Pilani -- Dubai Campus, Dubai International Academic City, Dubai, UAE 345055
\and
Department of Mechanical Engineering, Birla Institute of Technology and Science, Pilani -- Pilani Campus, Vidya Vihar, Pilani, India 333031\\
\email{adarsh@dubai.bits-pilani.ac.in}}

\maketitle              

\begin{center}
{\footnotesize
$^\dagger$ These authors contributed equally to this work\\
$^*$ Corresponding author
}
\end{center}

\begin{abstract}
We study the emergence of Cosmological Frequency Combs (CFCs) in a quintessence cosmology with an exponential potential using a dynamical systems formulation. Expressing the evolution equations in expansion-normalized variables yields an autonomous nonlinear system that supports time-periodic attractors corresponding to limit cycles, producing comb-like spectral structures in cosmological observables without external periodic forcing. Numerical simulations reveal transitions between single frequency, comb like and chaotic regimes controlled by the fundamental frequency, background equation of state parameter, and initial conditions. Coherent comb structures arise only within well defined dynamical windows, while very low frequencies and unfavorable initial conditions suppress phase locking. These results show that CFCs naturally emerge from nonlinear cosmological dynamics and motivate further study of their possible observational implications.
\keywords{Dark energy; scalar field dynamics; Limit cycles; Spectral structures; Late-time cosmic evolution}
\end{abstract}

\section{Introduction}
\justify
The discovery of the present day accelerated expansion of the Universe, first established through Type Ia supernova observations in the late 1990s \cite{SupernovaSearchTeam:1998fmf,Perlmutter:1998np}, remains one of the central turning points in modern cosmology. In general relativity the simplest explanation is a cosmological constant $\Lambda$ \cite{Weinberg:1988cp}, which together with cold dark matter gives the concordance $\Lambda$CDM model. Although this model fits a wide range of cosmological data with impressive accuracy, it still leaves important questions open, especially the cosmological constant problem and the persistent tension between early and late Universe determinations of the Hubble constant \cite{Martin:2012bt,Planck:2018vyg}. These issues have motivated many alternatives to a strictly constant dark-energy component, including modified gravity \cite{Nojiri:2017ncd}, dynamical scalar field models such as quintessence \cite{Odintsov:2023weg} and effective ideas inspired by quantum gravity settings, such as braneworld and loop-inspired cosmologies \cite{Sahni:2002dx}. Many of these models can reproduce late time acceleration, but they often do so by adding new parameters or by requiring some degree of tuning. It is therefore useful to look for dynamical mechanisms that can generate richer late time behavior without simply imposing it by hand.
\\
\\
Dynamical systems methods provide a natural way to study this question. By rewriting the cosmological field equations in terms of expansion-normalized variables, one can analyze the qualitative behavior of the system through fixed points, attractors, bifurcations, and limit-cycle dynamics, without needing exact analytic solutions \cite{dyn1Copeland:2006wr,dyn2bahamonde2018dynamical}. This approach has already been useful in understanding scalar field cosmologies, including scaling solutions, thawing and freezing behavior, and late-time attractors. In our earlier work, we introduced Cosmological Frequency Combs (CFCs) as a possible new dynamical phenomenon in cosmology \cite{trivedi2025cosmological}. The idea is motivated by the broader appearance of frequency-comb structures in nonlinear physical systems. Optical frequency combs are central to precision spectroscopy and metrology \cite{hall2006nobel,picque2019frequency}, while phononic systems can generate comb-like spectra through nonlinear three-wave mixing \cite{cao2014phononic,ganesan2017phononic,ganesan2018excitation,de2023mechanical}. Similar structures have also appeared in nonlinear magnonic systems \cite{wang2021magnonic,wang2022twisted}. These examples suggest that frequency combs are not tied to one particular physical platform, but can arise quite generally from nonlinear mode coupling and phase locking.
\\
\\
In the cosmological setting considered here, the relevant structure appears through stable, time-periodic attractors in the expansion-normalized phase space. These are nonlinear limit cycles rather than fixed points. Because the autonomous system governing exponential quintessence is intrinsically nonlinear, the scalar field variables can settle into self-sustained, phase-locked oscillations. These oscillations generate a hierarchy of harmonically related frequencies, which can then imprint comb-like structures onto cosmological observables such as the Hubble parameter and the growth rate of large-scale structure. Importantly, this behavior does not require an explicitly oscillatory potential or any external periodic driving; it arises from the nonlinear dynamics of the late-time cosmological system itself. Building on this framework, we numerically study the reduced autonomous system over broad ranges of the fundamental frequency and initial amplitudes. This allows us to map transitions between single frequency, comb like, and chaotic regimes, and to identify the dynamical windows in which coherent comb structures persist. We find that the onset and stability of the comb phase depend sensitively on the frequency scale and initial conditions, and that one sector can retain comb-like behavior even when the other begins to broaden or lose coherence. These results support the interpretation of CFCs as a genuine nonlinear feature of the cosmological dynamics, while also providing a basis for connecting them to possible late time observational signatures.
\\
\\
This paper is structured as follows. In Sction~2, we present the cosmological model and derive the autonomous dynamical system using expansion-normalized variables. In Section~3, we perform numerical simulations of the reduced system and analyze the resulting spectral behavior under variations of the fundamental frequency and initial conditions. Finally, Section~4 summarizes the main results and discusses their implications for nonlinear cosmological dynamics and possible observational signatures.

\section{Theoretical Basis of CFC}
\justify

We investigate a quintessence scalar field model minimally coupled to gravity. The theoretical framework is described by the action

\begin{equation}
S=\int d^{4}x\sqrt{-g}\left[\frac{R}{2\kappa^{2}}-\frac12 g^{\mu\nu}\partial_{\mu}\phi\partial_{\nu}\phi-V(\phi)\right].
\end{equation}

\justify
In this expression, $g_{\mu\nu}$ denotes the spacetime metric tensor and $R$ is the Ricci scalar curvature constructed from it. The parameter $\kappa^2 = 8\pi G$ represents the gravitational coupling constant, where $G$ is Newton’s gravitational constant. The field $\phi$ corresponds to a canonical scalar degree of freedom minimally coupled to gravity, while $V(\phi)$ specifies the scalar field potential governing the self-interaction of the field.

\justify
Assuming a spatially flat Friedmann–Lemaître–Robertson–Walker (FLRW) background geometry, the corresponding cosmological evolution equations are given by

\begin{align}
3H^{2} &= \kappa^{2}\!\left(\rho+\frac{\dot\phi^{2}}{2}+V\right), \\
2\dot H+3H^{2} &= -\kappa^{2}\!\left(w\rho+\frac{\dot\phi^{2}}{2}-V\right),\\
\ddot\phi+3H\dot\phi+V_{\phi}&=0,
\end{align}

\justify
Here, $\rho$ denotes the energy density of a background barotropic fluid characterized by a constant equation of state (EOS) parameter $w$. The Hubble expansion rate is defined as $H=\dot a/a$, where $a(t)$ is the cosmological scale factor and overdots denote derivatives with respect to cosmic time $t$. The scalar field $\phi$ acts as a dynamical dark-energy component that evolves under the influence of its potential $V(\phi)$ and contributes to the overall expansion of the Universe. The parameter $\kappa^2 = 8\pi G$ specifies the strength of the gravitational coupling.

\justify
Following Refs.~\cite{dyn1Copeland:2006wr,dyn2bahamonde2018dynamical,dyn3clifton2012modified}, we introduce the expansion-normalized variables

\begin{equation}
x\equiv\frac{\kappa\dot\phi}{\sqrt6\,H}, 
\qquad 
y\equiv\frac{\kappa\sqrt{V}}{\sqrt3\,H}.
\end{equation}

\justify
These dimensionless variables represent the relative contributions of the scalar field kinetic and potential energies to the total expansion rate of the Universe.

\justify
For the scalar field self-interaction, we consider an exponential potential of the form

\begin{equation}
V(\phi)=V_{0}e^{-\lambda\kappa\phi},
\qquad 
\lambda=\text{const}.
\end{equation}

\justify
Here, the constant parameter $\lambda$ determines the steepness of the potential and therefore controls the dynamical behavior of the scalar field.

\justify
With these definitions, the cosmological equations can be recast as an autonomous dynamical system for the variables $(x,y)$,

\begin{align}
x^\prime &= -\frac{3}{2} \left[ 2 x + (w-1) x^3 + x (w+1) (y^2 - 1) - \sqrt{\frac{2}{3}} \lambda y^2 \right], \\
y^\prime &= -\frac{3}{2} y \left[ (w-1) x^2 + (w+1) (y^2 - 1) + \sqrt{\frac{2}{3}} \lambda x \right].
\end{align}

\justify
For convenience in the subsequent analysis, the above equations may be expressed in the equivalent rearranged form

\begin{align}
x^\prime &= \frac{3}{2} (w-1) \left( x - x^3 \right) + \sqrt{\frac{3}{2}} y^2 \left( \lambda - x \sqrt{\frac{3}{2}} (w+1) \right), \\
y^\prime &= \frac{3}{2} (w+1) \left( y - y^3 \right) - \sqrt{\frac{3}{2}} x y \left( \lambda + x \sqrt{\frac{3}{2}} (w-1) \right).
\end{align}

\justify
To explore the possibility of oscillatory dynamics within this nonlinear system, we represent the variables $x$ and $y$ using harmonic expansions of the form

\begin{align}
x &= \frac{1}{2} \left( u e^{i \omega N} + u^* e^{-i \omega N} \right), \\
y &= \frac{1}{2} \left( v e^{i \frac{\omega}{2} N} + v^* e^{-i \frac{\omega}{2} N} \right),
\end{align}

\justify
where $u$ and $v$ denote complex amplitudes and $\omega$ represents the fundamental oscillation frequency. The variable $N=\ln a$ corresponds to the number of e-folds and serves as the time variable commonly used in the dynamical systems analyses of cosmology.

\justify
The derivatives and nonlinear combinations appearing in the equations are given by

\begin{align}
x^\prime &= \frac{1}{2} \left( i \omega u e^{i \omega N} + u^\prime e^{i \omega N} - i \omega u^* e^{-i \omega N} + {u^\prime}^* e^{-i \omega N} \right), \\
y^\prime &= \frac{1}{2} \left( i \frac{\omega}{2} v e^{i \frac{\omega}{2} N} + v^\prime e^{i \frac{\omega}{2} N} - i \frac{\omega}{2} v^* e^{-i \frac{\omega}{2} N} + {v^\prime}^* e^{-i \frac{\omega}{2} N} \right), \\
y^2 &= \frac{1}{4} \left( v^2 e^{i \omega N} + v^{*2} e^{-i \omega N} + 2 |v|^2 \right), \\
xy &= \frac{1}{4} \left( u v e^{\frac{3 i \omega}{2} N} + u v^* e^{\frac{i \omega}{2} N} + u^* v e^{-\frac{i \omega}{2} N} + u^* v^* e^{-\frac{3 i \omega}{2} N} \right), \\
x^3 &= \frac{1}{8} \left( u^3 e^{3 i \omega N} + 3 |u|^2 u e^{i \omega N} + 3 |u|^2 u^* e^{-i \omega N} + u^{*3} e^{-3 i \omega N} \right), \\
y^3 &= \frac{1}{8} \left( v^3 e^{\frac{3 i \omega}{2} N} + 3 |v|^2 v e^{\frac{i \omega}{2} N} + 3 |v|^2 v^* e^{-\frac{i \omega}{2} N} + v^{*3} e^{-\frac{3 i \omega}{2} N} \right).
\end{align}

\justify
Substituting these expressions into the dynamical equations and retaining only the dominant harmonic contributions leads to the following amplitude equations

\begin{align}
(i \omega u + u^\prime) &= \frac{3}{2} (w-1) u + \frac{1}{2} \sqrt{\frac{3}{2}} \lambda v^2, \\
\left( i \frac{\omega}{2} v + v^\prime \right) &= \frac{3}{2} (w+1) v - \frac{1}{2} \sqrt{\frac{3}{2}} \lambda u v^*.
\end{align}

\justify
Defining

\begin{equation} \label{omegas}
\Omega_u = \frac{3}{2}(w-1), 
\qquad 
\Omega_v = \frac{3}{2}(w+1),
\qquad 
\alpha = \frac{1}{2} \sqrt{\frac{3}{2}} \lambda,
\end{equation}

\justify
the system can be written in the compact form

\begin{align}
u^\prime &= (\Omega_u - i \omega) u + \alpha v^2, \\
v^\prime &= \left( \Omega_v - i \frac{\omega}{2} \right) v - \alpha u v^*.
\end{align}
\section{Numerical Simulations}

\begin{figure}
    \centering
    \includegraphics[width=\linewidth]{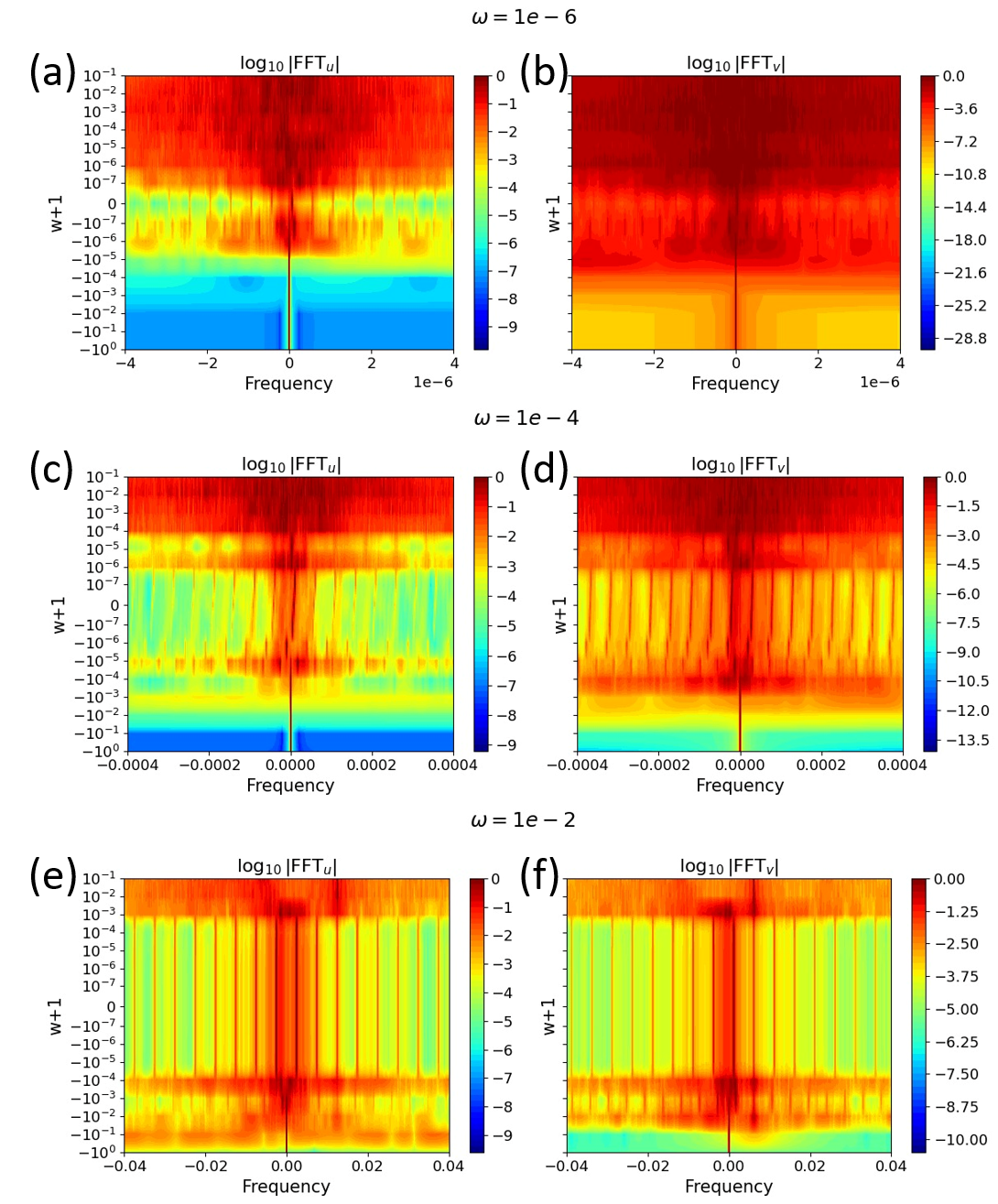}
    \caption{Fourier spectra of the numerical solutions illustrating the dependence on the fundamental frequency $\omega$. Panels (a,c,e) display $\log_{10}|\mathrm{FFT}_u|$, while panels (b,d,f) show $\log_{10}|\mathrm{FFT}_v|$ for $\omega = 10^{-6},\,10^{-4},\,10^{-2}$, respectively. The spectra are shown as functions of the Fourier frequency and the shifted EOS parameter $(w+1)$ within the interval $[-10^{-7},\,10^{-7}]$. The simulations are performed with $\Omega = 5\times10^{-1}$, $\Omega/2 = 2.5\times10^{-1}$, and $\alpha = \frac{1}{2}\sqrt{\frac{3}{2}}\,\lambda$.}
    \label{fig1}
\end{figure}

\begin{figure}
    \centering
    \includegraphics[width=\linewidth]{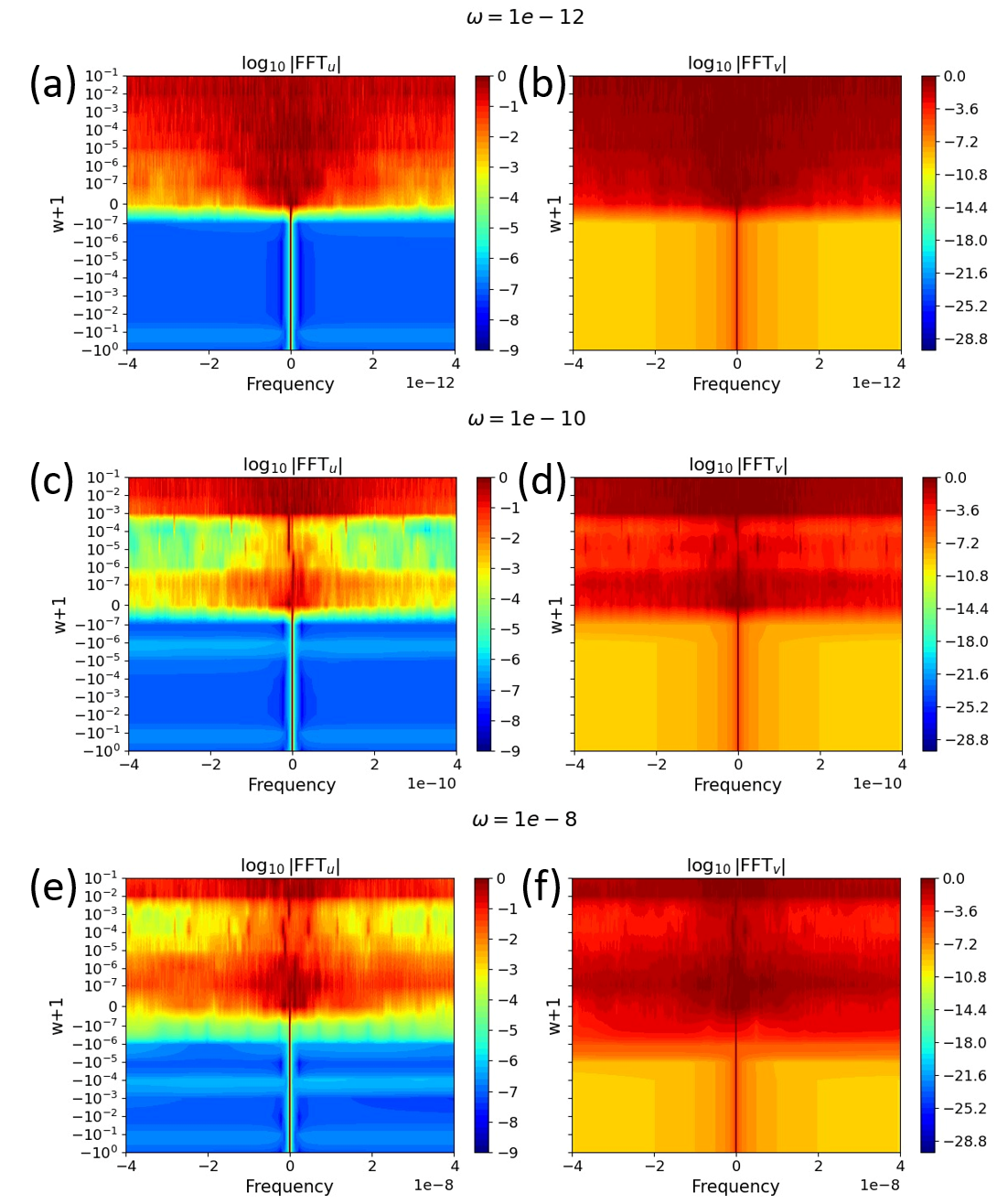}
    \caption{Fourier spectra at ultra low fundamental frequencies. Panels (a,c,e) correspond to $\log_{10}|\mathrm{FFT}_u|$, and panels (b,d,f) to $\log_{10}|\mathrm{FFT}_v|$ for $\omega = 10^{-12},\,10^{-10},\,10^{-8}$, respectively. The spectral amplitudes are shown as functions of the Fourier frequency and the shifted EOS parameter $(w+1)$ over the range $[-10^{-7},\,10^{-7}]$. All simulations use $\Omega = 5\times10^{-1}$, $\Omega/2 = 2.5\times10^{-1}$, and $\alpha = \frac{1}{2}\sqrt{\frac{3}{2}}\,\lambda$.}
    \label{fig1a}
\end{figure}

\begin{figure}
    \centering
    \includegraphics[width=\linewidth]{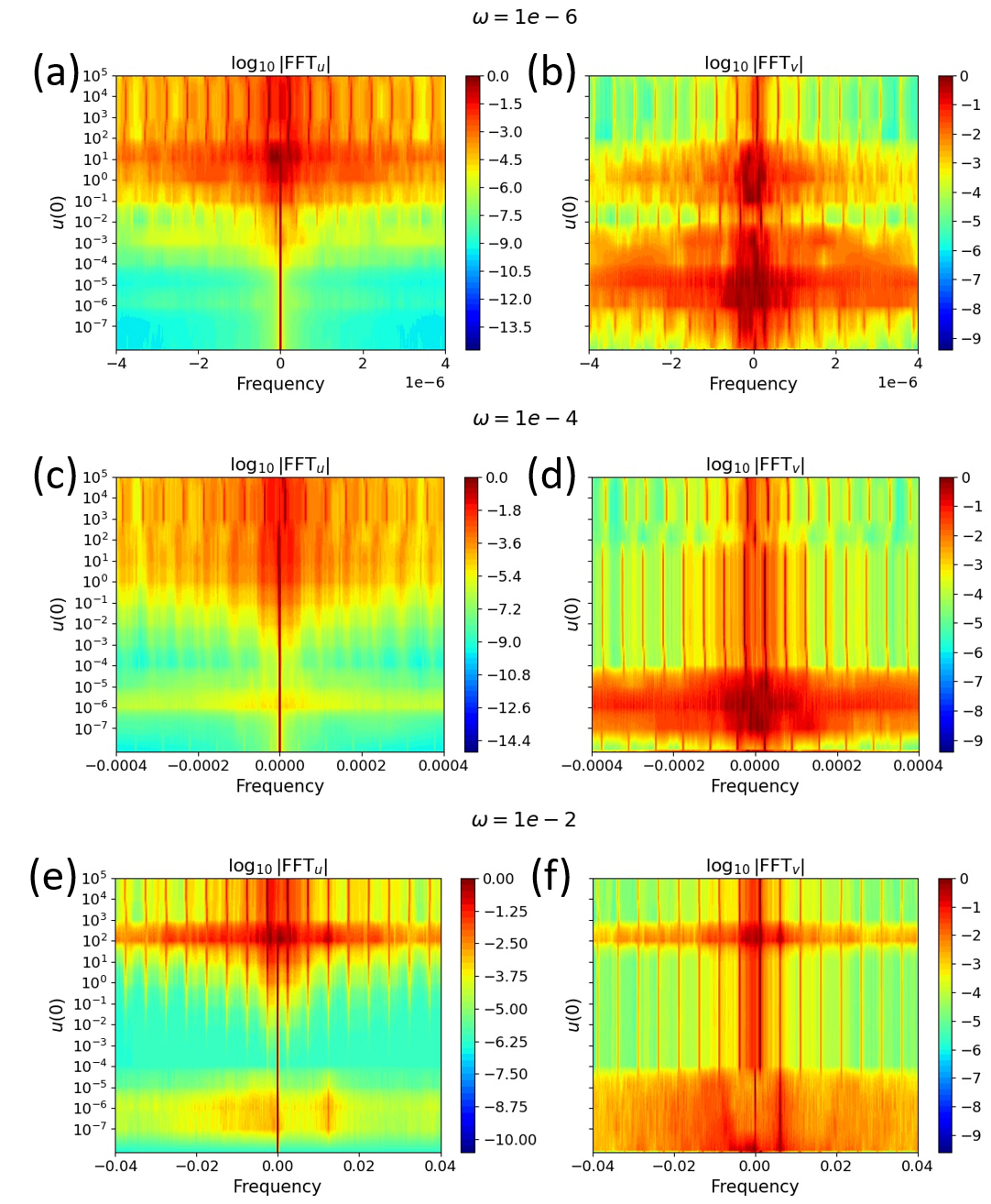}
    \caption{Dependence of the frequency spectra on the initial condition $u(0)$ for moderate and high fundamental frequencies. Panels (a,c,e) show $\log_{10}|\mathrm{FFT}_u|$, while panels (b,d,f) show $\log_{10}|\mathrm{FFT}_v|$ for $\omega = 10^{-6},\,10^{-4},\,10^{-2}$, respectively. The spectra are presented as functions of the Fourier frequency and the initial amplitude $u(0)$ spanning $[10^{-10},\,10^{5}]$. The remaining parameters are fixed at $\Omega = 5\times10^{-1}$, $\Omega/2 = 2.5\times10^{-1}$, and $\alpha = \frac{1}{2}\sqrt{\frac{3}{2}}\,\lambda$.}
    \label{fig2}
\end{figure}

\begin{figure}
    \centering
    \includegraphics[width=\linewidth]{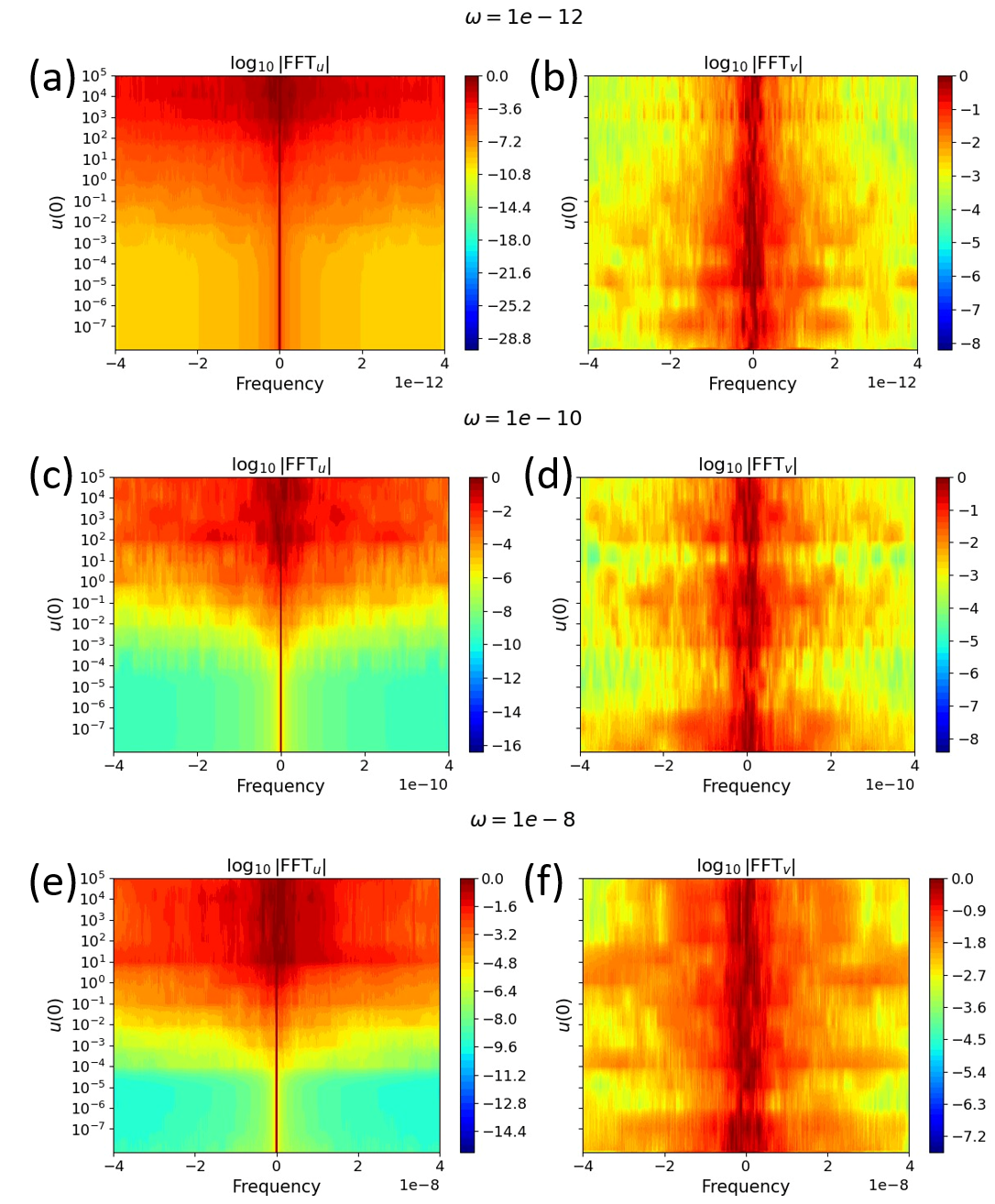}
    \caption{Frequency spectra as functions of the initial condition $u(0)$ for ultra low fundamental frequencies. Panels (a,c,e) correspond to $\log_{10}|\mathrm{FFT}_u|$, and panels (b,d,f) to $\log_{10}|\mathrm{FFT}_v|$ for $\omega = 10^{-12},\,10^{-10},\,10^{-8}$, respectively. The spectral response is shown over the Fourier frequency domain and initial amplitudes $u(0)\in[10^{-10},\,10^{5}]$. The simulations are carried out with $\Omega = 5\times10^{-1}$, $\Omega/2 = 2.5\times10^{-1}$, and $\alpha = \frac{1}{2}\sqrt{\frac{3}{2}}\,\lambda$.}
    \label{fig2b}
\end{figure}

\begin{figure}
    \centering
    \includegraphics[width=\linewidth]{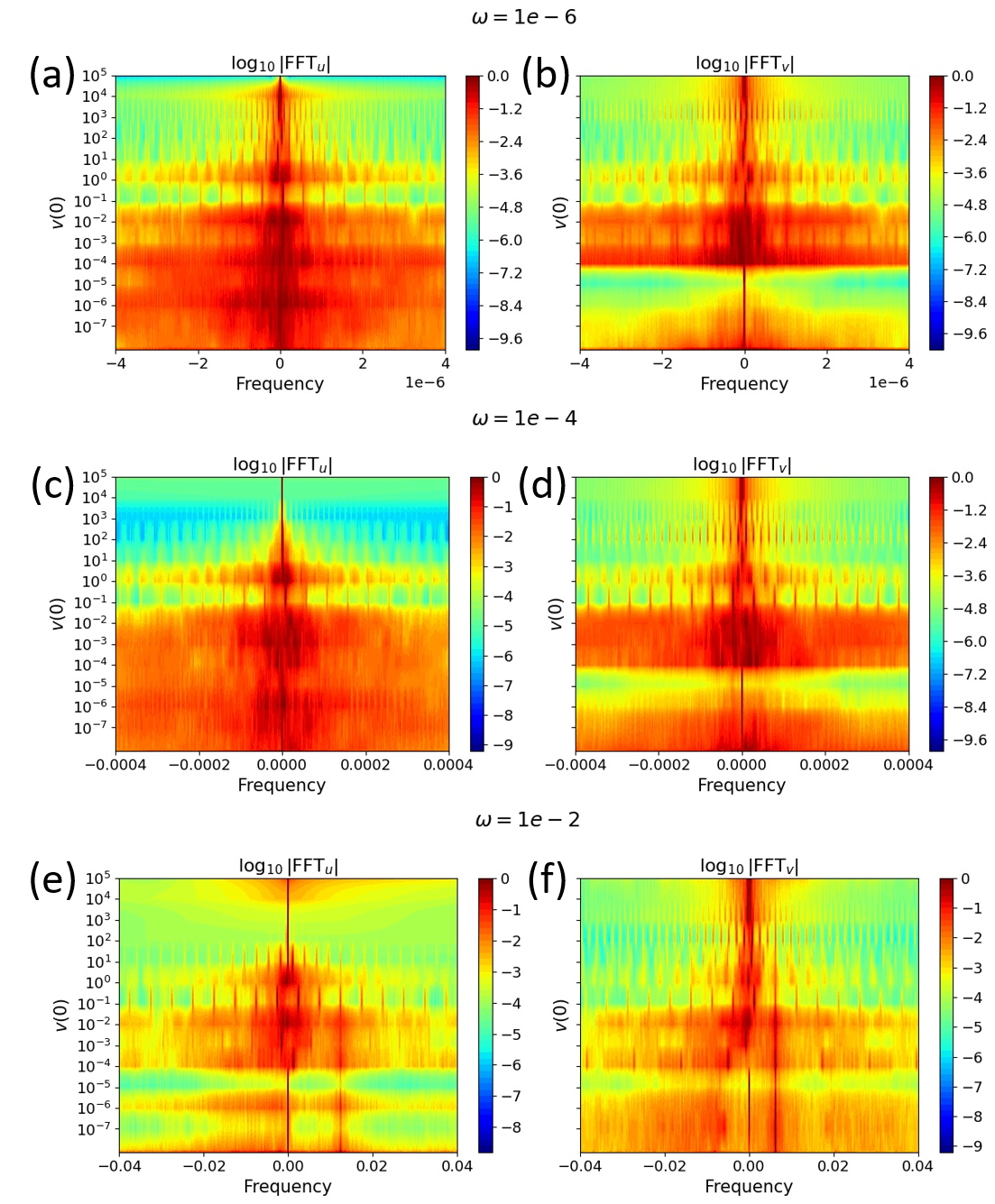}
    \caption{Frequency spectra illustrating the dependence on the initial condition $v(0)$ for higher fundamental frequencies. Panels (a,c,e) show $\log_{10}|\mathrm{FFT}_u|$, while panels (b,d,f) show $\log_{10}|\mathrm{FFT}_v|$ for $\omega = 10^{-6},\,10^{-4},\,10^{-2}$, respectively. The spectra are displayed as functions of the Fourier frequency and the initial amplitude $v(0)$ in the range $[10^{-10},\,10^{5}]$, with $\Omega = 5\times10^{-1}$, $\Omega/2 = 2.5\times10^{-1}$, and $\alpha = \frac{1}{2}\sqrt{\frac{3}{2}}\,\lambda$.}
    \label{fig3}
\end{figure}

\begin{figure}
    \centering
    \includegraphics[width=\linewidth]{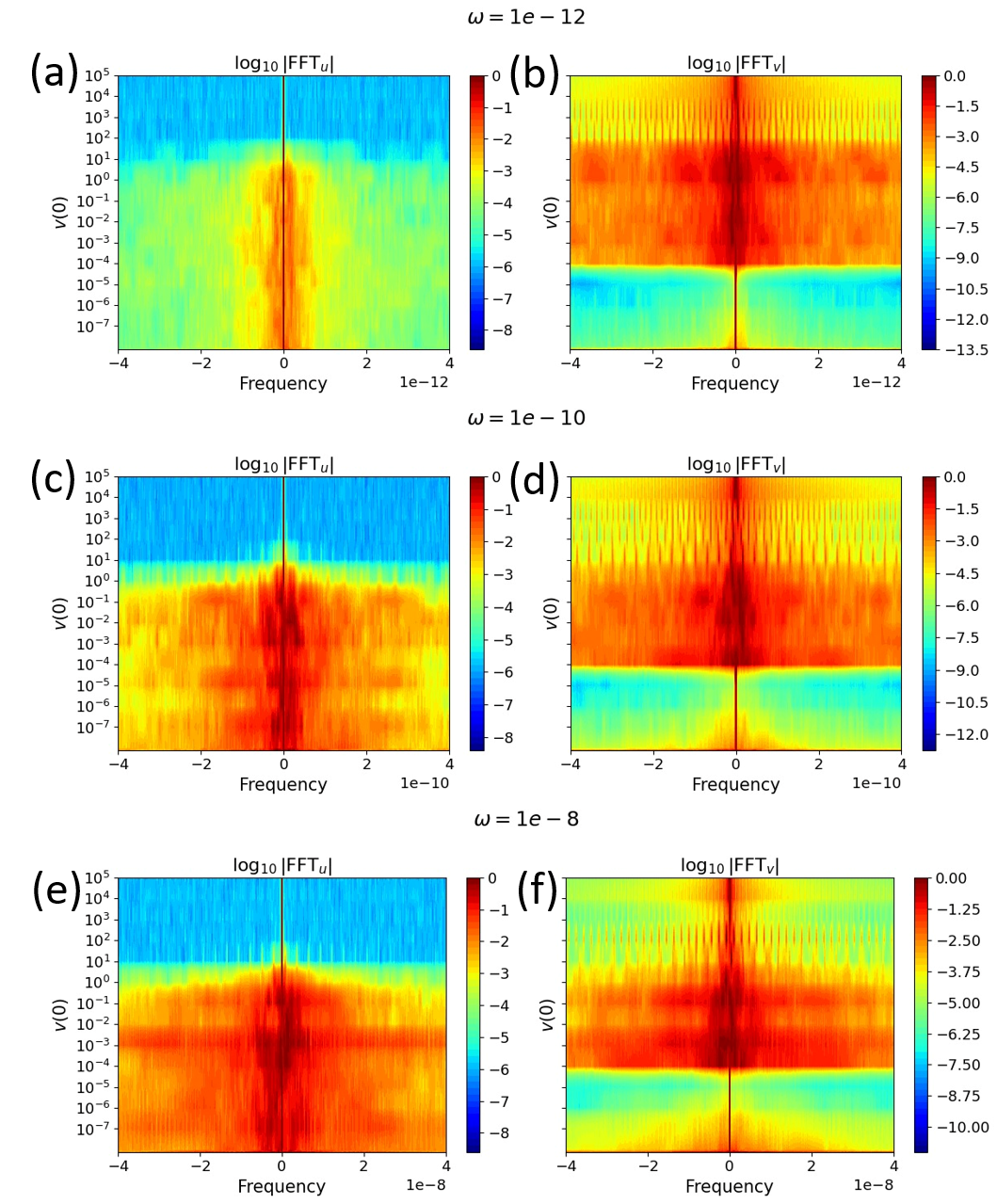}
    \caption{Frequency spectra as functions of the initial condition $v(0)$ for ultra low fundamental frequencies. Panels (a,c,e) correspond to $\log_{10}|\mathrm{FFT}_u|$, and panels (b,d,f) to $\log_{10}|\mathrm{FFT}_v|$ for $\omega = 10^{-12},\,10^{-10},\,10^{-8}$, respectively. The spectral amplitudes are shown across the Fourier frequency domain and initial amplitudes $v(0)\in[10^{-10},\,10^{5}]$. The remaining parameters are fixed at $\Omega = 5\times10^{-1}$, $\Omega/2 = 2.5\times10^{-1}$, and $\alpha = \frac{1}{2}\sqrt{\frac{3}{2}}\,\lambda$.}
    \label{fig3b}
\end{figure}

\justify  
We numerically integrate Eqs.~(22) and (23) under varying initial conditions $u(0)$, $v(0)$, and the EOS parameter $w$ to systematically explore the dynamical regimes accessible in the system. This parameter sweep allows us to identify how nonlinear interactions redistribute spectral energy and drive transitions between ordered and disordered motion. The resulting spectra, presented in Figs.~\ref{fig1}--\ref{fig3b}, reveal three characteristic behaviors: (i) a single frequency regime in which the spectral power is sharply localized around the fundamental mode, (ii) a coherent comb regime characterized by a set of regularly spaced harmonics with preserved phase coherence, and (iii) a broadband chaotic regime marked by strong spectral spreading and loss of harmonic organization. These regimes correspond to successive stages in the nonlinear destabilization of regular oscillatory motion as either the intrinsic frequency scale or the effective nonlinearity of the system is increased.
\justify
Fig.~\ref{fig1} displays the spectral response as a function of the shifted EOS parameter $(w+1)$ for different values of the fundamental frequency $\omega$, the system evolves from a nearly single frequency state to a comb regime and, for sufficiently large $(w+1)$, to chaotic dynamics. For small $(w+1)$, nonlinear contributions remain weak and the spectral power is concentrated near the fundamental oscillation mode. As $(w+1)$ increases, nonlinear coupling becomes more effective, generating higher harmonics and producing a regular comb-like structure in the spectrum. These comb structures are most prominent for smaller values of $\omega$, where well defined harmonic bands persist across extended parameter ranges, indicating that low-frequency oscillations allow coherent nonlinear energy transfer to develop gradually. However, with increasing $\omega$, these coherent structures weaken and the system transitions to a chaotic regime.

\justify
Figure~\ref{fig1a} shows that the ultra low frequency regime behaves somewhat differently from the higher-frequency cases. For $\omega = 10^{-12}$, the spectra remain strongly localized near zero frequency over a wide range of $(w+1)$, which indicates a mostly single frequency response with very weak mode coupling. Once $(w+1)$ crosses a threshold value, however, the system moves directly into an irregular, chaotic regime rather than passing through a broad, well developed comb phase. The same pattern appears for larger ultra low frequencies, but the transition happens earlier wherein for $\omega = 10^{-8}$, the onset of chaotic behavior occurs at lower values of $(w+1)$ than for $\omega = 10^{-12}$.
\justify

Figure~\ref{fig2} shows how the spectra depend on the initial condition $u(0)$ for different values of the fundamental frequency $\omega$. For small amplitudes, $u(0)\lesssim10^{-4}$, the response is essentially linear, with the spectral power concentrated around the fundamental frequency. As $u(0)$ is increased, nonlinear effects excite higher harmonics, which organize into a regular banded structure and produce a coherent comb regime over the approximate range $10^{-4}\lesssim u(0)\lesssim10^{1}$. At still larger amplitudes, this harmonic organization begins to break down and the spectra broaden, signaling the transition toward chaotic dynamics. The size of the comb window depends on $\omega$ wherein lower frequencies allow coherent comb formation over a wider range of $u(0)$, while higher frequencies tend to push the system into chaos at smaller initial amplitudes. It is also notable that comb like structures can remain visible in the $v$ spectra even when the corresponding $u$ spectra have already broadened substantially, showing that the two sectors do not lose coherence in exactly the same way.
\justify

Figure~\ref{fig2b} gives the corresponding $u(0)$ dependence in the ultra low frequency regime. In this case the spectra stay sharply localized near the fundamental mode across almost the entire range, indicating a robust single frequency state. Only weak and narrowly spaced chaotic features appear in a limited intermediate interval, approximately $10^{-2}\lesssim u(0)\lesssim10^{1}$, and these do not grow into a fully broadband chaotic spectrum. Figures \ref{fig3} and \ref{fig3b} then show the dependence on the initial condition $v(0)$. Unlike the $u(0)$ case, coherent comb structures are most prominent for relatively lower initial values, roughly $10^{-4}\lesssim v(0)\lesssim10^{-1}$, where the system is excited enough to support nonlinear phase locking but not so strongly that it immediately enters strong mode mixing. For very small $v(0)$, the spectra become irregular because the excitation is too weak to sustain stable harmonic organization. As $v(0)$ moves outside the intermediate range and the effective nonlinear coupling becomes stronger, the spectra broaden progressively, with fully broadband chaotic behavior appearing for $v(0)\gtrsim10^{1}$. As $\omega$ increases, the comb window shifts toward smaller values of $v(0)$.
\\

\section{Subtleties of CFC dynamics}

The discussion of the CFC above studied the reduced system numerically and identified single frequency, comb like and chaotic regimes as functions of $\omega$, $(w+1)$, $u(0)$, and $v(0)$. In the original CFC work these oscillations were interpreted as modulations of $H(N)$, growth and lensing observables and so it is useful to make the connection between the reduced dynamical variables and the cosmological observables more explicit. In particular, one would like to know when the dynamical windows found in the reduced system are simply useful features of the nonlinear toy dynamics and when they correspond to physically meaningful regions of late time cosmology. This also helps clarify how far the reduced amplitude equations can be trusted as a description of the original cosmological variables. We therefore construct a small amplitude bridge between the harmonic CFC variables and observable quantities and then use it to isolate the physically relevant part of the CFC regime. We begin by treating the CFC oscillations as small corrections around a smooth late time scalar field background. Instead of assuming that the normalized variables $x$ and $y$ are purely oscillatory, we write
\begin{equation}
x(N)=x_0+X(N),
\end{equation}
\begin{equation}
y(N)=y_0+Y(N),
\end{equation}
where $x_0$ and $y_0$ describe the slowly varying background configuration, while $X(N)$ and $Y(N)$ contain the oscillatory CFC correction. Over the relatively short interval on which the CFC modulation is followed, the background quantities can be treated as approximately constant and the leading harmonic corrections may then be written as
\begin{equation}
X(N)=\frac{1}{2}\left(Ue^{i\omega N}+U^{*}e^{-i\omega N}\right),
\end{equation}
\begin{equation}
Y(N)=\frac{1}{2}\left(Ve^{i\omega N/2}+V^{*}e^{-i\omega N/2}\right),
\end{equation}
where $U$ and $V$ are slowly varying complex amplitudes associated with the two sectors of the reduced system. This notation keeps the background dark-energy configuration separate from the CFC envelope, and makes it possible to translate numerical windows in $u(0)$ and $v(0)$ into corresponding windows for physical observables. The scalar field density fraction is $\Omega_\phi=x^2+y^2$ and substituting the decompositions above gives
\begin{equation}
\Omega_\phi(N)=\left(x_0+X\right)^2+\left(y_0+Y\right)^2
\end{equation}
Expanding to leading order in the oscillatory amplitudes gives
\begin{equation}
\Omega_\phi(N)=\Omega_{\phi 0}+2x_0X(N)+2y_0Y(N)+X^2(N)+Y^2(N)
\end{equation}
where
\begin{equation}
\Omega_{\phi 0}=x_0^2+y_0^2
\end{equation}
In the small amplitude regime, $|X|,|Y|\ll 1$, the quadratic terms are subleading, and the leading modulation of the scalar field fraction becomes
\begin{equation}
\delta\Omega_\phi(N)\simeq 2x_0X(N)+2y_0Y(N).
\end{equation}
This gives the first analytic link between the reduced CFC variables and cosmological observables. It also shows that the observable modulation is not determined by $U$ and $V$ alone, but by how those oscillatory amplitudes project onto the background scalar field configuration through $x_0$ and $y_0$. 
\\
\\
The same bridge can be used to estimate the modulation of the Hubble rate and for this, we perturb the Friedmann equation around a smooth background solution and write the CFC induced fractional correction to the Hubble rate as
\begin{equation}
H(N)=\bar H(N)\left[1+\epsilon_H(N)\right]
\end{equation}
one obtains, to leading order
\begin{equation}
\epsilon_H(N)\simeq \frac{1}{2}\delta\Omega_\phi(N).
\end{equation}
Using the expression for $\delta\Omega_\phi(N)$, this gives
\begin{equation}
\epsilon_H(N)\simeq x_0X(N)+y_0Y(N).
\end{equation}
Therefore the characteristic amplitude of the Hubble modulation is approximately
\begin{equation}
\epsilon_H^{\rm amp}\simeq x_0|U|+y_0|V|.
\end{equation}
For a late time dark energy configuration close to $w_\phi=-1$, the kinetic contribution is small and the potential contribution dominates. In this regime one typically has $x_0\ll y_0$, with $y_0\simeq \sqrt{\Omega_{{\rm DE},0}}$ and taking $\Omega_{{\rm DE},0}\simeq 0.7$ gives
\begin{equation}
y_0\simeq \sqrt{0.7}\simeq 0.84
\end{equation}
Thus, in the dark energy dominated limit the Hubble modulation is approximately
\begin{equation}
\epsilon_H^{\rm amp}\simeq 0.84 |V|.
\end{equation}
A few percent modulation of the late time Hubble rate, of the size needed to create a noticeable shift between a local and background value of $H_0$, therefore corresponds to
\begin{equation}
|V|\sim 0.04-0.06
\end{equation}
This gives the first important consequence of the analytic bridge. The numerical CFC regime found in the $v$ sector, especially the intermediate range where coherent comb structures appear before strong broadband mixing sets in, naturally overlaps with the amplitude range required for a percent level modulation of $H(N)$. The $v$-dominated CFC window is therefore not only a useful feature of the reduced equations. and in a late time dark energy configuration close to $w_\phi=-1$, it can also correspond to a physically meaningful modulation scale. The same analysis constrains the physically viable part of the $u$ sector. The scalar field equation of state is
\begin{equation}
w_\phi=\frac{x^2-y^2}{x^2+y^2}.
\end{equation}
Therefore
\begin{equation}
1+w_\phi=\frac{2x^2}{x^2+y^2}.
\end{equation}
Using $x=x_0+X$ and averaging over the fast oscillatory phase gives
\begin{equation}
\left\langle x^2\right\rangle=x_0^2+\left\langle X^2\right\rangle.
\end{equation}
For the harmonic form of $X(N)$, the averaged quadratic correction is of order $|U|^2/2$, and hence
\begin{equation}
\left\langle 1+w_\phi\right\rangle\simeq \frac{2x_0^2+|U|^2}{\Omega_{\phi 0}},
\end{equation}
up to corrections higher order in the oscillation amplitudes. If late time observations require the scalar field to remain close to $w_\phi=-1$, then the oscillatory kinetic amplitude must satisfy
\begin{equation}
|U|^2\lesssim \Omega_{\phi 0}\Delta w,
\end{equation}
where $\Delta w$ denotes the allowed departure from $w_\phi=-1$. For example, if $\Delta w\lesssim 0.05$ and $\Omega_{\phi 0}\simeq 0.7$, then
\begin{equation}
|U|\lesssim \sqrt{0.7\times 0.05}\simeq 0.19.
\end{equation}
Thus very large values of $u(0)$, although useful for mapping the nonlinear transition to chaos, cannot be interpreted directly as realistic late time dark energy amplitudes without further qualification. A more physical late time CFC window is the small amplitude part of the broader numerical range, roughly where $|U|\ll 1$ and preferably $|U|\lesssim 10^{-1}$.
\\
\\
A second point concerns the validity of the reduced amplitude equations themselves, and for this we can write the reduced CFC system is
\begin{equation}
U'=(\Omega_u-i\omega)U+\alpha V^2,
\end{equation}
\begin{equation}
V'=\left(\Omega_v-\frac{i\omega}{2}\right)V-\alpha UV^*,
\end{equation}
where we have the definitions as in \eqref{omegas}. These equations arise after retaining the dominant resonant harmonic contributions and neglecting higher harmonics and nonresonant nonlinear terms. This reduction is controlled only when the amplitudes remain small,
\begin{equation}
|U|^2+|V|^2\ll 1
\end{equation}
The nonlinear terms retained in the amplitude system must also remain perturbative relative to the leading harmonic components and this requires conditions of the schematic form
\begin{equation}
\alpha |V|^2\ll |U|,
\end{equation}
\begin{equation}
\alpha |U||V|\ll |V|,
\end{equation}
\begin{equation}
|\Omega_u||U|^3\ll |U|,
\end{equation}
\begin{equation}
|\Omega_v||V|^3\ll |V|.
\end{equation}
Equivalently, the cubic self interactions and higher order mixing terms must remain subdominant compared with the leading resonant exchange between the $U$ and $V$ sectors. These inequalities tell us why the large amplitude chaotic regimes are valuable for understanding the nonlinear phase structure of the reduced system, but should not automatically be identified with viable late time cosmologies.
\\
\\
A further condition comes from averaging over nonresonant harmonics as well. The harmonic truncation assumes that nonresonant terms oscillate rapidly enough to average away over the timescale on which the amplitudes evolve and a useful dimensionless measure of this requirement is
\begin{equation}
\mathcal{R}_{\rm nr}= \frac{
\max\left(
\alpha |U|,\alpha |V|,
|\Omega_u|(|U|^2+|V|^2),
|\Omega_v|(|U|^2+|V|^2)\right)
}{\omega}
\end{equation}
A clean resonant amplitude description requires
\begin{equation}
\mathcal{R}_{\rm nr}\ll 1.
\end{equation}
If this condition is not satisfied, the reduced equations may still be useful as a qualitative model of nonlinear phase locking but the interpretation of the solution as a sharply resolved frequency comb becomes less straightforward. This point matters especially in the ultra low frequency regime and when $\omega$ is very small, the nonresonant pieces do not necessarily average away over the cosmological time interval accessible to observations. This leads to the second important physical conclusion, wherein the small $\omega$ regime, CFCs may realistically appear as slow late time offsets or residual drifts in $H(N)$, growth, or lensing, rather than as directly resolved combs in low-redshift data. The model can still produce a phase coherent modulation capable of shifting locally inferred cosmological quantities, but its observational signature would look almost quasi-static over the redshift range probed by current late-time surveys. A genuinely resolved cosmological frequency comb would require at least one full oscillatory cycle across the available interval in $N$.
\\
\\
This can be stated quantitatively by counting the number of CFC cycles over a redshift window as since $N=\ln a$, the interval between redshift zero and a maximum redshift $z_{\rm max}$ is
\begin{equation}
\Delta N=\ln(1+z_{\rm max}).
\end{equation}
The number of oscillatory cycles across this interval is then
\begin{equation}
N_{\rm cyc}=\frac{\omega \Delta N}{2\pi}.
\end{equation}
For late time data extending to $z_{\rm max}\simeq 2$, one has
\begin{equation}
\Delta N\simeq \ln 3\simeq 1.1.
\end{equation}
For $\omega=10^{-2}$ this gives
\begin{equation}
N_{\rm cyc}\simeq \frac{10^{-2}\times 1.1}{2\pi}\simeq 1.7\times 10^{-3}.
\end{equation}
This is far below one complete cycle and so, over the observable late time range $z\lesssim 2$, such a modulation would show up as a slowly varying drift or offset not as a resolved sequence of comb teeth. To obtain even one full cycle over the same interval, one would need
\begin{equation}
\omega\gtrsim \frac{2\pi}{\Delta N}\simeq 6.
\end{equation}
Several resolved teeth would require still larger values of $\omega$ and that regime lies outside the small $\omega$ approximation used in the basic analytical construction and would need a separate analysis of the full system without relying on the same harmonic truncation. These points also make clearer how CFC dynamics differs from more standard nonlinear behavior in cosmology. The main novelty of CFC is not just that they generate another nonlinear attractor, scaling solution, or oscillatory correction but rather it is that they can produce phase-coherent, harmonically related residuals across several observables. The leading Hubble modulation derived above implies that distance observables, growth observables and lensing observables should inherit correlated residual structures with related phases. One should therefore not look only for a generic departure from $\Lambda$CDM, but for a structured residual pattern tied to the same underlying oscillatory phase. This can be illustrated at the level of the growth equation as well, as the matter perturbation equation in $N$ is
\begin{equation}
\delta''+\left(2+\frac{H'}{H}\right)\delta'-\frac{3}{2}\Omega_m(N)\delta=0.
\end{equation}
Writing
\begin{equation}
H(N)=\bar H(N)\left[1+\epsilon_H\cos(\omega N+\varphi)\right]
\end{equation}
one obtains, to leading order
\begin{equation}
\frac{H'}{H}= \frac{\bar H'}{\bar H}
-\epsilon_H\omega \sin(\omega N+\varphi).
\end{equation}
The matter fraction also changes because $\Omega_m\propto H^{-2}$ at fixed matter density, so
\begin{equation}
\Omega_m(N)\simeq \bar\Omega_m(N)\left[1-2\epsilon_H\cos(\omega N+\varphi)\right]
\end{equation}
If the growth rate is written as
\begin{equation}
f(N)=\frac{d\ln \delta}{dN},
\end{equation}
then the growth equation becomes
\begin{equation}
f'+f^2+\left(2+\frac{H'}{H}\right)f-\frac{3}{2}\Omega_m=0.
\end{equation}
Now we can write
\begin{equation}
f(N)=\bar f(N)+\Delta f(N),
\end{equation}
where $\bar f(N)$ satisfies the smooth background equation and keeping only terms linear in $\Delta f$ and $\epsilon_H$ gives
\begin{equation}
\Delta f' +
\left[
2\bar f+2+\frac{\bar H'}{\bar H}
\right]\Delta f =  \bar f,\epsilon_H\omega \sin(\omega N+\varphi) 3\bar\Omega_m\epsilon_H\cos(\omega N+\varphi).
\end{equation}
This expression shows that the same CFC phase that modulates $H(N)$ also produces a locked response in the growth sector. The growth residual is not arbitrary too and it is forced by the same cosine and sine structure that appears in the expansion history. This phase correlation is the main observational feature that can distinguish a genuine CFC signal from a generic dynamical dark energy residual. Putting these estimates together, the physically relevant CFC region is the small amplitude, near-$w=-1$, $v$ dominated part of the nonlinear window. In this region, values of $|V|\sim 0.04-0.06$ can naturally corresponding to a few percent Hubble modulation while keeping the scalar field configuration close to dark energy domination. But very large values of $|U|$ or $|V|$ are better viewed as probes of the mathematical transition to broadband chaos, rather than as immediately viable late time cosmologies and also, ultra low values of $\omega$ are best interpreted as producing quasi-static offsets or slow residual drifts across the observable redshift range, while a directly resolved comb requires larger frequencies and a treatment beyond the small $\omega$ approximation. Please note that this does not weaken the narrative of CFC but rather, they identify the part of the nonlinear dynamical landscape most relevant for cosmology and clarify what kind of observational signature should be expected. 
\section{Conclusion}
\justify
In this work, we have shown that Cosmological Frequency Combs can emerge as a natural consequence of nonlinear dynamics in a simple quintessence cosmology with an exponential potential. By expressing the cosmological evolution equations in expansion-normalized variables, the system reduces to an autonomous nonlinear dynamical system that admits stable time periodic solutions in the form of limit cycles, rather than only fixed point attractors. These limit cycles generate sets of harmonically related frequencies, producing comb like spectral structures in cosmological observables without the need for externally imposed oscillatory potentials or explicit time dependence. Our numerical analysis has shown that the existence and stability of the comb regime are restricted to well defined dynamical windows controlled by the fundamental frequency, the background EOS parameter, and the initial conditions of the system. Higher fundamental frequencies favor coherent phase locking and sustained harmonic generation, while lower frequencies enhance nonlinear mixing and lead more rapidly to broadband, chaotic spectra while at extremely low frequencies, nonlinear effects become too weak to support comb formation and the dynamics approach a predominantly single frequency state. The initial conditions further determine clear transitions between single frequency, comb like, and chaotic regimes, and in some cases comb structures persist in one dynamical variable even when they are suppressed in the other, reflecting asymmetric nonlinear coupling. We also clarified the exact regime in which the CFC results provide a cosmologically interesting window, and when they could best be interpreted as pedagogical. Together, these results clarify how nonlinear mode coupling and phase locking shape late time cosmological dynamics and provide a foundation for future studies of the stability, generality, and possible observational signatures of Cosmological Frequency Combs in more realistic cosmological settings.
\\
\\
\section*{Acknowledgements}
The work of O.T. is supported in part by the Vanderbilt Discovery Doctoral Fellowship.

\end{document}